# Passive Optical Network Approach to GigaHertz-Clocked Multiuser Quantum Key Distribution

Veronica Fernandez *Member, IEEE*, Robert J. Collins *Member, IEEE*, Karen J. Gordon, Paul D. Townsend, and Gerald S. Buller *Member, IEEE*

*Abstract*— We present the application of quantum key distribution technologies to fiber-based broadband passive optical access networks. This application is based on our 850 nm wavelength gigahertz clock-rate single-receiver system, is compatible with existing telecommunications fiber and exploits a wavelength band not currently utilized in access networks. The developed quantum key distribution networks are capable of transmitting over distances consistent with the span of access links for metropolitan networks (10 km), at clock frequencies ranging up to 3 GHz.

## I. INTRODUCTION

QUANTUM KEY DISTRIBUTION (QKD) [1] exploits fundamental physical principles to achieve information security on optical communications links. To date, much of the work on fiber-based QKD has focused on extending the achievable transmission distance on point-to-point (P2P) links to beyond 100 km [2]. Operation over this length scale would open the possibility of applications in secure, metropolitan-area-sized networks. However, comparatively less attention has been given to the question of how QKD could best be implemented on the access links to such networks. If based on current network topologies and technologies then these access links are likely to have a span of up to around 10 km and to be based on either multiple P2P links or point-to-multipoint passive optical networks (PONs). Optical access solutions of these types are currently in deployment in a number of regions of the world, where network operators are upgrading pre-existing copper-based access networks with optical fiber in order to supply new, high bandwidth services to customers. The two main fiber-based replacement architectures are shown in stylized form in Fig. 1. For simplicity, only the downstream part of the network is shown with the transmitter(s) (Tx) at the central node and the receivers (Rx) at the end-user locations. However, in practice each end-user would also contain a transmitter to enable upstream communication to a receiver in the central node. Wavelength division multiplexing (WDM) is usually employed to allow the two channels to share the same fiber without interference, with the upstream channel typically operating in the wavelength band around 1300 nm and the downstream channel (or channels) operating in the wavelength band around 1500 nm [3]. In the case of the PON, the passive optical splitter is shown as a single "*star*", but other architectures with, for example, distributed splitters in "*tree-*" or "*bus-*" type configurations are also possible. In this paper we address the question of how QKD could be best implemented in a secure network application employing these typical access topologies and wavelength allocation plans.

We have previously developed a *single-receiver* QKD system using the B92 protocol [4] with polarization encoding that is capable of transmission at gigahertz clock rates through the use of mature silicon single-photon avalanche diodes (Si-SPADs) and 850 nm light in standard telecommunications fiber [5][6]. Here we demonstrate the feasibility of extending the application of this system to multi-user access networks. In this scenario the 850 nm operating wavelength brings a

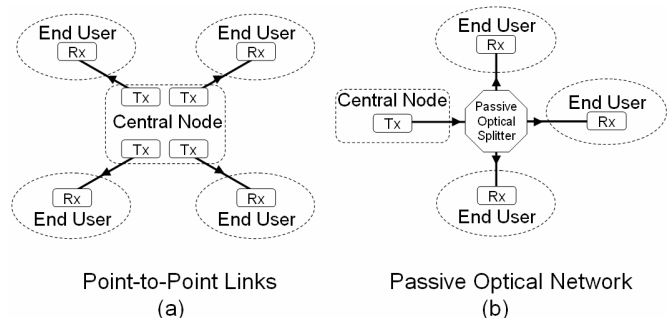

Fig. 1. The two main architectures for broadband fiber-access networks.(a) shows a point-to-point architecture with the transmitters (Tx) in the center and receivers (Rx) at each arm. (b) shows a point-to-multipoint passive optical network architecture where the transmitter is connected to each receiver via a passive optical splitter.

Manuscript received August 11, 2006; revised October 11, 2006. This work was supported in part by the European Commission SECOQC Integrated Project and United Kingdom Engineering and Physical Sciences Research Council (project reference GR/N12466). Paul Townsend would like to thank Science Foundation Ireland for support under grant number 03/IN1/1340.

V. Fernandez, R. J. Collins , K. J. Gordon, and G. S. Buller are with the School of Engineering and Physical Sciences, Heriot-Watt University, Edinburgh, EH14 AS, UK (e-mail: G.S.Buller@hw.ac.uk).

P. D. Townsend is with the Photonics Systems Group, Tyndall National Institute, University College Cork, Lee Maltings, Prospect Row, Cork, Ireland.

Digital Object Identifier 10.1109/JQE.2006.887175

number of potential advantages. For example, the use of both 1300 nm and 1500 nm wavelength bands for conventional channels in access networks makes wavelength space relatively congested. The 850 nm band is unused and is widely separated from the high intensity conventional channels making cross-talk suppression via optical filtering relatively straightforward. In addition the relatively short target distance ($\leq 10$ km) ensures that the quantum channel fiber loss (~2.2 dBkm$^{-1}$) is not prohibitively high. These advantages are qualified by the need to operate conventional fibers and components at a non-standard wavelength. However, our work shows that this need not be a limiting factor. We present results obtained from the experimental implementation of two gigahertz clocked, multi-user, polarization based, B92 protocol QKD systems and provide an analysis of the fraction of the final key shared by two or more receivers when *weak coherent pulses* (WCP) are used to simulate true single-photon states. In addition, we show that although the polarization dependent loss (PDL) for a telecommunications wavelength splitter operated at a wavelength of 850 nm must be considered, QKD is still possible. We note that the B92 protocol is particularly vulnerable to the "intercept-resend" eavesdropping attack with a lossless channel if there is significant channel loss. However, B92 is relatively simple to implement in practice, and the protocol is suitable for experimental tests of the implementation of QKD in PONs. Most importantly, the main conclusions reached in this work are applicable to less vulnerable protocols, such as BB84 [1].

## II. SINGLE-RECEIVER QUANTUM KEY DISTRIBUTION

### A. Net Bit Rate

The *net bit rate* ($R(\Delta t)_{net}$) is defined as the bit rate per unit time after error correction [7] and privacy amplification [8] have been applied to distil the bits received by Bob into a usable cryptographic key. For our measurements, this was estimated using [9][10]:

$$R(\Delta t)_{net} \approx \left[1 + Q\log_2(Q) - \frac{7}{2}Q - I_{AE}\left(1 - (1-Q)\log_2(1-Q) - \frac{7}{2}Q\right)\right] \cdot R(\Delta t)_{sifted} \quad (1)$$

where $Q$ is the *quantum bit error rate* (QBER) [11], $R(\Delta t)_{sifted}$ is the bit-rate after temporal filtering [5], and $I_{AE}$ is the maximum information shared between Alice and Eve (the eavesdropper), defined by:

$$I_{AE(\max)} = 1 - \cos\theta \quad (2)$$

for the B92 protocol, where $\theta$ is the relative angle between the quantum states.

In our calculations of $R(\Delta t)_{net}$ we assume a scenario in which Eve use a technique which allows her to unambiguously extract the maximum information from the transmission between Alice and Bob. Therefore, for our separation angle of 45°, the maximum information that could be gained by Eve is approximately 29% [12].

### B. Single-Receiver System

The single-receiver system is shown in Fig. 2. In this system the sender (Alice) uses a photon source with a wavelength of 850 nm to allow the receiver (Bob) to take full advantage of the mature technology of Si-SPADs, as used previously in QKD applications in both free-space [13] and in specialist 850nm single-mode fiber links [14]. The most appropriate single-photon detection technology at $\lambda \sim 1550$ nm, InGaAs/InP SPADs [15][16] exhibit the deleterious effects of after-pulsing which place severe restrictions on the maximum possible count rate [2].

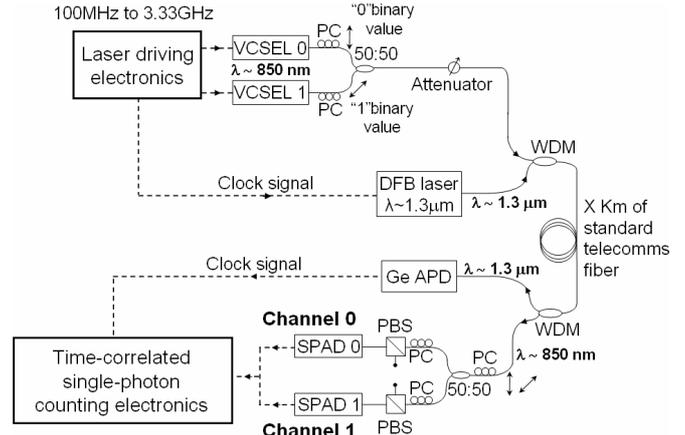

Fig. 2. Schematic diagram of the single-receiver quantum key distribution system. This system was to form the basis of the later P2P and point-to-multipoint PON experiments. PBS: Polarization splitter. PC: Polarization controller. WDM: Wavelength division multiplexer. APD: Avalanche photodiode. VCSEL: Vertical-cavity surface-emitting laser. SPAD: Single-photon avalanche diode. DFB: Distributed feedback laser. 50:50: An equal ratio beam-splitter/coupler. The double headed arrows represent the polarization states of individual photons propagating in the optical fiber.

The quantum channel is composed of 9 μm diameter-core standard telecommunications fiber which is single-mode at $\lambda \sim 1.3$ μm/1.55 μm. In standard telecommunications optical fiber, the 850 nm wavelength light propagates as two modes due to mode dispersion which would lead to an increase in the QBER. Constructing both Alice and Bob from 5 μm diameter-core fiber, which allows only a single mode to propagate at a wavelength of 850 nm, and fusion splicing a short (<1 m) length of 5 μm diameter-core fiber on the ends of the 9 μm diameter-core telecomms fiber prevents the propagation of the second order LP$_{11}$ mode [17]. Using this technique, more than 99% of the photons are launched into the fundamental (LP$_{01}$) mode. This allows Alice and Bob to use 850 nm wavelength light and silicon single-photon detectors in conjunction with the same standard telecommunications fiber used for the longer wavelength channels.

To ensure security, a QKD system must operate at the single-photon level. Our system highly attenuates the bright laser pulses from the vertical-cavity surface-emitting lasers (VCSELs) to a conventional mean number of photons per pulse ($\mu$) of 0.1 immediately prior to entering the main transmission medium, the telecommunications fiber.

Using a modified Si-SPAD module [18][19] the single-receiver system has been extensively tested at clock rates from 1 to 2 GHz [5][6], significantly higher than systems using InGaAs/InP single-photon detectors operating at a wavelength of 1.55 μm [2].

There are three primary factors that contribute to the QBER of the QKD system; the polarization leak of the components

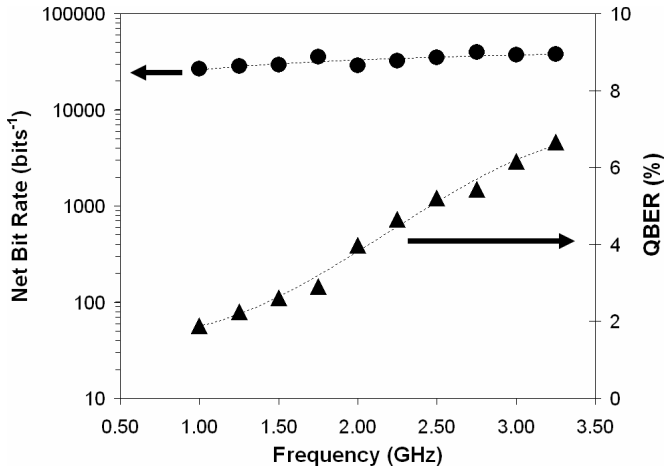

Fig. 3. Net bit rate and QBER at different clock frequencies for a fixed fiber length of 6.55 km in the single-receiver QKD system.

($R(\Delta t)_{leak}$), the dark counts of the detector ($R(\Delta t)_{dark}$) and the temporal response of the VCSELs and the SPADs ($R(\Delta t)_v$). These contributions can be represented in the following manner:

$$QBER(\Delta t) = \frac{R(\Delta t)_{leak} + R(\Delta t)_{dark} + R(\Delta t)_v}{R(\Delta t)_{sifted}}. \quad (3)$$

We have now adapted and optimized our driving electronics to increase the maximum operating frequency of the single-receiver system to 3.25 GHz, as shown in Fig. 3. These increased bit transmission rates proportionally improve the bit-rate at each receiver in a multi-user system, as will be seen later.

### III. MULTI-USER QUANTUM KEY DISTRIBUTION

The high bit-rates achieved for the single-receiver QKD by this group presented earlier [5][6] provided the motivation for the implementation of multi-user QKD applications [20][21] where one transmitter can share a different and unique quantum key with any of the multiple receivers on the network. This form of network communication is *one-to-one communication* since the multiple Bobs do not share a key between themselves and each develops their own unique key with Alice – any secret communications between each Bob must pass through Alice.

#### A. PON Approach to Multiple P2P Links

In a P2P link, as shown in Fig. 1(a), there is a separate transmitter for each receiver. However, it is possible to share one set of Alice optics and electronics between all the end users by placing a passive optical splitter within the central node (see Fig. 4). Since Alice was constructed using optical components that are single-mode at $\lambda \sim 850$ nm, a fused biconical taper (FBT) technology 1×8 optical splitter, which is also single-mode at $\lambda \sim 850$ nm, was used to divide the optical signal from Alice into the separate fiber links to the 8 receivers. As in the single-receiver system, the transmission channel was composed of standard telecommunication fiber, with fusion splices made between the 5 µm diameter-core fiber and the 9 µm diameter-core fiber at points B and C in Fig. 4 to prevent the propagation of the $LP_{11}$ mode.

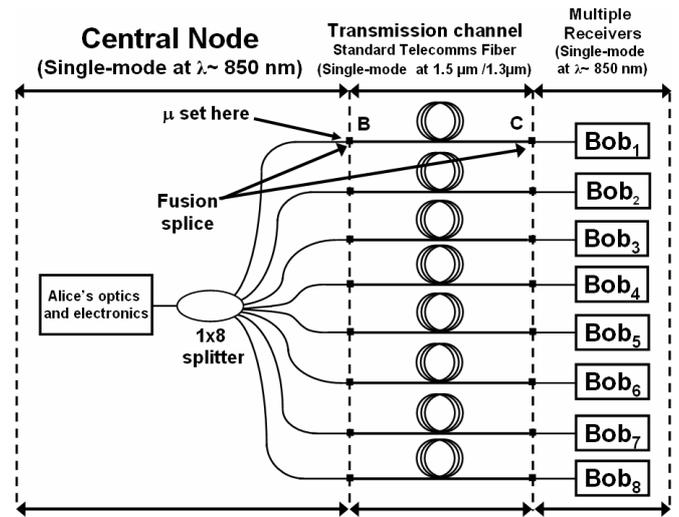

Fig. 4. The PON approach to multiple point-to-point links showing the passive optical splitter within the central node. Alice, the splitter and all of the Bobs are constructed from fiber which is single-mode at a wavelength of 850 nm. The transmission channels between Alice and the Bobs are standard telecomms fiber.

WDMs can be used at both the central node and each of the end users to allow the 850 nm-wavelength quantum communications channel to be transmitted without interference on the same fiber as the classical communications channels using a method analogous to the optical clocking shown in Fig. 2.

This multi-user architecture (which we shall call PON P2P) was analyzed using two different mean photon numbers per laser pulse ($\mu$) – a conventional $\mu$ of 0.1 photons per pulse at each output arm and an aggregate $\mu$ of 0.1 over all 8 arms. Setting $\mu$ after the splitter allows for the compensation of the internal loss of the optical splitter offering the possibility of higher bit-rates.

The first mean photon number was chosen to ensure that the optical path between point *B* and each Bob (see Fig. 4) was equivalent to a single-receiver QKD system. We note, of course, that this particular approach gives an additional security disadvantage because a powerful eavesdropper with the capability to monitor all 8 output ports simultaneously would receive an aggregate mean photon number per pulse of up to ~ 0.8. This is why the second case with an aggregate $\mu$ of 0.1 was also analyzed for comparison.

#### B. Experimental Results for PON P2P Multi-User Architecture

In this section we report a selection of results taken using a non-return to zero (NRZ) differential output from a pre-programmed pulse pattern generator (PPG) operating at a clock frequency of 1.25 GHz to drive the VCSELs. We report the net bit rate and the measured QBER at four different output ports of the $\lambda \sim 850$ nm optical splitter — spliced to different lengths of standard telecomms fiber. The data for each output port (representing a different receiver) was taken sequentially. Fig. 5 shows the net bit rate and QBER plotted against fiber length for a clock frequency of 1.25 GHz.

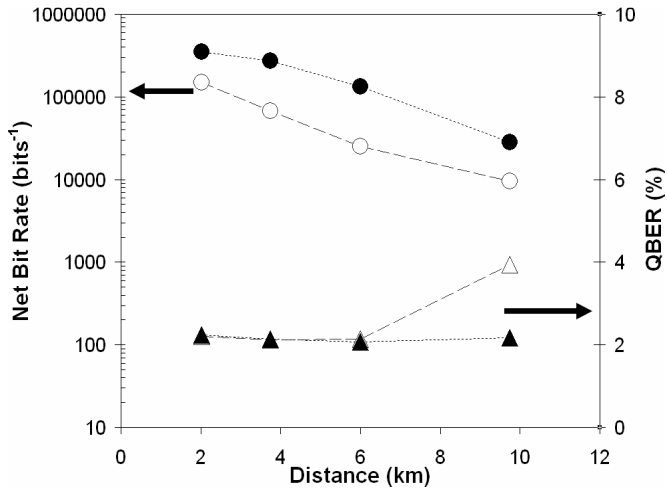

Fig. 5. Net bit rate and QBER versus fiber length for a clock frequency of 1.25 GHz for the PON P2P multi-user architecture. The filled points were obtained using a photon number $\mu$ of 0.1 photons per pulse at each of the output arms of the 1×8 splitter. The outline points represent an aggregate $\mu$ of 0.1 photons per pulse at the 8 outputs of the splitter. Each distance in the graph represents a different receiver.

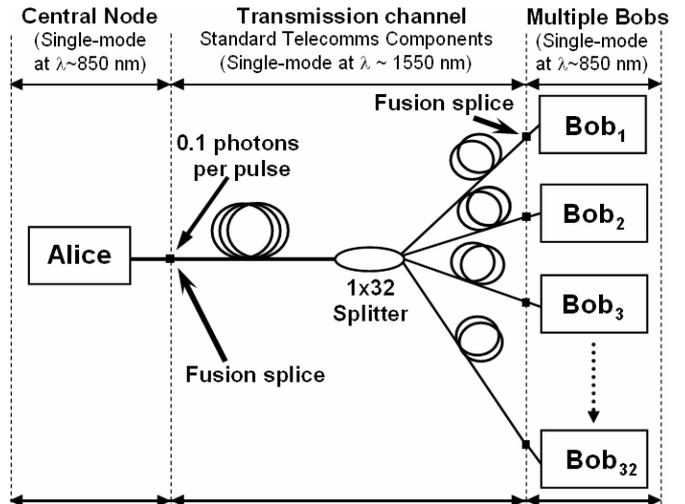

Fig. 7. The QKD point-to-multipoint PON. Alice and all of the Bobs are constructed from fiber which is single-mode at a wavelength of 850 nm, the quantum channel is constructed from fiber which is single-mode at a wavelength of 1.55 µm.

The QBER is almost constant over the measured fiber length range with an increase at the longest fiber length analyzed. This increase is due to the relatively higher contribution to the QBER caused by dark counts generated by each SPAD ($R(\Delta t)_{dark}$) with respect to $R(\Delta t)_{sifted}$ in (3). As the transmission distance is increased, $R(\Delta t)_{sifted}$ decreases as opposed to the probability of a dark event occurring in a time window, $\Delta t$, which remains constant. This in turn means that as the transmission distance is increased, the contribution to the QBER caused by the dark counts is comparatively higher.

We also report the net bit rate and QBER at different clock frequencies for one receiver at a fixed fiber length of 6.01 km. Fig. 6 shows the net bit-rate and the QBER at increasing clock frequencies.

As the clock frequency is increased, the bit width is reduced, which causes an increase in the contribution to the QBER introduced by the temporal response of both the VCSELs and the SPADs, $R(\Delta t)_V$. This is due to the relatively slower temporal response of the VCSELs and SPADs with respect to the bit width, causing the detection times of arriving photons to spread into adjacent time windows. In this case, the increasing QBER with clock frequency causes $R(\Delta t)_{net}$ to show only a marginal increase with clock frequency, as detailed in Equation (1). However, if lower timing jitter detectors are used then a more substantial increase in $R(\Delta t)_{net}$ with clock frequency could be achieved.

### C. Point-to-Multipoint PON Multi-User Architecture

The point-to-multipoint PON uses a 1×32 standard telecommunications optical splitter (fabricated using ion-exchange technology and single-mode at a wavelength of $\lambda \sim 1550$ nm) within the insecure transmission channel (see Fig. 7). The input and output ports of the optical splitter were fusion spliced to the single-mode at $\lambda \sim 850$ nm optical fiber from Alice and Bob respectively, as in the previous QKD multi-user configuration, to suppress the propagation of the $LP_{11}$ mode within the splitter.

### D. Experimental Results for the Point-to-Multipoint PON

In this section we report a selection of results taken at a clock frequency of 1 GHz. As in the previous case, the lasers were driven using a NRZ differential output from a pre-programmed PPG. We report the net bit rate and the measured QBER at four different output ports of the 1×32 telecomms wavelength optical splitter spliced to different lengths of standard telecomms fiber. The data for each output port (representing a different receiver) was taken sequentially. Fig. 8 shows the net bit rate and QBER (%) plotted against fiber length for each Bob.

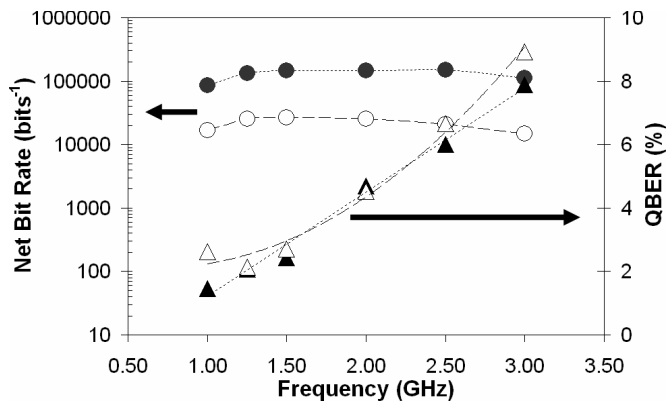

Fig. 6. Net bit rate and QBER versus clock frequency for a fixed fiber length of 6.01 km for one user of the PON P2P multi-user architecture. The filled points were obtained using a photon number $\mu$ of 0.1 photons per pulse at each of the output arms of the 1×8 splitter. The outline points represent an aggregate $\mu$ of 0.1 photons per pulse at the 8 outputs of the splitter.

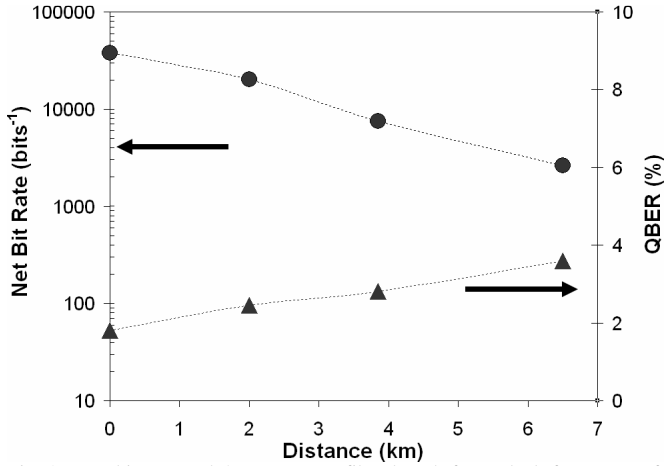

Fig. 8. Net bit rate and QBER versus fiber length for a clock frequency of 1 GHz for the point-to-multipoint PON. Each distance in the graph represents a different receiver.

We also report the net bit rate and QBER at different clock frequencies for one receiver at a fixed fiber length of 6.01 km. Fig. 9 shows the net bit-rate and the QBER at increasing frequencies.

### E. Consequences of PDL in the Point-to-Multipoint PON

The polarization dependent loss (PDL) of an optical component is defined as the variation in the transmitted power as the polarization angle of the input light is rotated through all the possible states of polarization [22][23]. The insertion loss for a particular output port of a splitter is the loss between the input and the output port, which depends on the splitting ratio, and any imperfections of the device. These imperfections are commonly referred to as the excess loss and include the average value of the maximum and minimum loss caused by the PDL.

The PDL and insertion loss were measured for both splitters at a wavelength of $\lambda \sim 850$ nm. The average insertion loss for each output port of the $\lambda \sim 850$ nm optical coupler/splitter was measured to be 13.28 dB, compared to the value of 10.00 dB

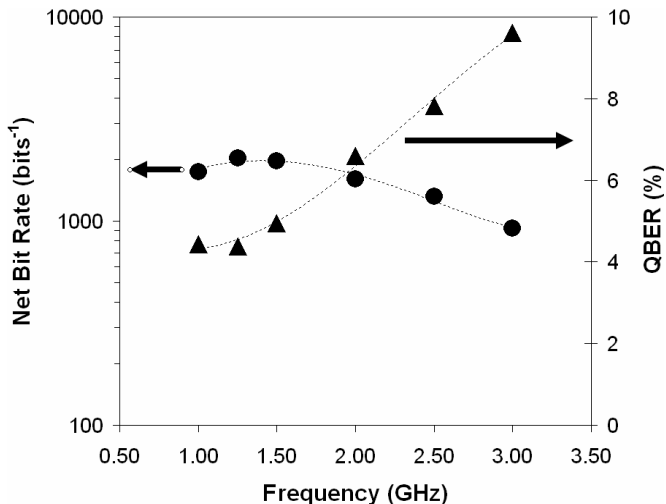

Fig. 9. Net bit rate and QBER versus clock frequency for a fixed fiber length of 6.01 km for one receiver of the point-to-multipoint PON.

quoted by the manufacturer. The measured PDL was less than 0.6 dB for each port.

The PDL and insertion loss of the telecomms wavelength splitter were also characterized. The average insertion loss for each port was measured to be 17.99 dB, compared to the value of 16.81 dB quoted by the manufacturer for $\lambda \sim 1550$ nm. However, significant differences in the insertion losses of some ports compared to others were observed. The maximum difference between ports was measured to be 9.52 dB. The implications of this will be analyzed in *Section G*. The PDL measured for the telecomms splitter was also greater than the values quoted by the manufacturer for $\lambda \sim 1550$ nm. At the design wavelength the splitter had a PDL of less than 0.4 dB, compared to a maximum measured value of 2.23 dB at $\lambda \sim 850$ nm.

The *PDL axis* is defined as the axis of the material for which the attenuation is at a minimum when the polarization state of the launched light is parallel to this axis [23]. The effect of the PDL on the linearly polarized non-orthogonal states of the B92 protocol is a different attenuation on the components of the electric fields of the "1"s or "0"s parallel and perpendicular to the PDL axis, which provokes a variation in the relative angle, $\alpha$, between the states. A theoretical analysis was developed to calculate this variation in $\alpha$ and in the intensity of the electric fields of the "1"s and "0"s for a given PDL. We should note that in a real world-scenario the linearly polarized states from Alice are likely to evolve into elliptical states prior to entering the splitter, as a consequence of the birefringence induced mainly by the optical fiber transmission medium. The theoretical analysis was also considered for this more general case and identical results were obtained in terms of the variation in the relative angle $\alpha$ and in the intensity of the electric fields induced by the PDL of the splitter.

Let us assume that $a$ and $b$ are the angles of the electric field of the linearly polarized states encoding "1"s and "0"s ($\vec{E}_1$ and $\vec{E}_0$) with the horizontal $x$ axis when they enter the splitter (see Fig. 10) and $a'$ and $b'$ are the angles of the electric field of the linearly polarized states encoding "1"s and "0"s ($\vec{E}_1{}'$ and $\vec{E}_0{}'$) with the horizontal $x$ axis when they exit the splitter. Then $a'$ and $b'$ can be expressed as a function of $a$ and $b$ of the form:

$$a' = \tan^{-1}(\tan(a) \times 10^{\frac{|PDL|}{20}})$$

$$b' = \tan^{-1}(\tan(b) \times 10^{\frac{|PDL|}{20}}), \qquad (4)$$

where $0° \leq a \leq 90°$ and $0° \leq b \leq 90°$ and the PDL is expressed in dB. (4) is still valid in the remaining quadrants, provided 180º is added to the values obtained for $a'$ and $b'$ in the second and third quadrants and 360° for the fourth quadrant. Therefore, the relative angle of the non-orthogonal states when they exit the splitter, $\alpha'$, is given by:

$$\alpha' = |a' - b'|. \qquad (5)$$

From (4) and (5) it can clearly be seen that the relative angle between the non-orthogonal states experiences a periodic variation that is dependent on the angles $a$ and $b$.

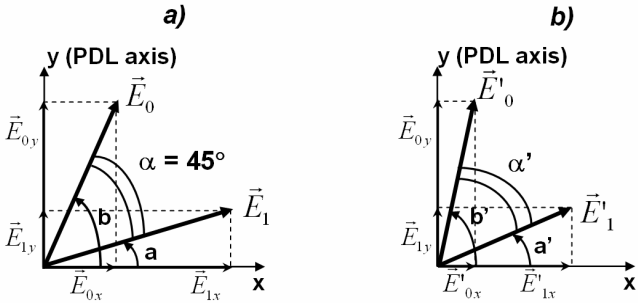

Fig. 10. The effect of PDL on the components of the electric field. (a) shows the state of the electric field vectors before entering the component with PDL, while (b) shows these vectors after propagating through the component with PDL.

As a consequence of the attenuation of the components of the electric fields $\vec{E}_1$ and $\vec{E}_0$ parallel to the $x$ axis due to the PDL, the intensity of the electric fields when they exit the component with PDL ($|\vec{E}_1'|^2$ and $|\vec{E}_0'|^2$) are given by:

$$|\vec{E}_0'|^2 = \frac{\cos^2 b}{\cos^2 b'}|\vec{E}_0|^2 10^{-\frac{|PDL|}{10}}$$

$$|\vec{E}_1'|^2 = \frac{\cos^2 a}{\cos^2 a'}|\vec{E}_1|^2 10^{-\frac{|PDL|}{10}}. \quad (6)$$

The effect on $R(\Delta t)_{net}$ of the maximum measured PDL of 2.23 dB for the point-to-multipoint PON is shown in Fig. 11. $R(\Delta t)_{net}$ was calculated using (1) for the highest QBER measured against transmission distance for the point-to-multipoint PON (3.59%), as shown in Fig. 8.

For a system with no PDL $\alpha$ remains constant as both states propagate through the transmission channel from Alice to each Bob ($\alpha = \alpha' = 45°$). $R(\Delta t)_{net}$ for this case is represented in Fig. 11 with a filled circle. However, when a PDL of 2.23 dB is present in the system, the original $R(\Delta t)_{net}$ for $\alpha = 45°$ is decreased by an amount that depends on the angles $a$ and $b$, and hence on $\alpha'$. This decrease (shown by the hollow circles in Fig. 11) is more noticeable for high values of $\alpha'$, which are a consequence of a high attenuation experienced by the states. The maximum attenuation of the electric field components occurs when the incident angles $a$ and $b$ are such that the electric field components parallel to the x-axis are maximal. This was the case for incident angles $a = -22.5°$ and $b = 22.5°$ or $a = 157.5°$ and $b = 202.5°$, which gave rise to the maximum separation angle, $\alpha_{MAX}' = 56°$ for a PDL of 2.23 dB.

For the calculation of $R(\Delta t)_{net}$ in Fig. 11 we assumed the worst-case scenario in which Eve intercepts the string of photons from Alice after the splitter and performs an unambiguous state discrimination measurement [12]. Hence the information she can obtain is given by (2) with $\theta = \alpha'$, where $\alpha'$ could vary between $\alpha'_{MIN} = 36°$ and $\alpha'_{MAX} = 56°$ for a PDL of 2.23 dB. We assumed that Bob is performing non-optimized state discrimination (as in our system), and hence the information he can obtain is given by:

$$R_{sifted} = \frac{1}{2}(1 - \cos^2 \alpha')(|\vec{E}_0'|^2 + |\vec{E}_1'|^2), \quad (7)$$

where $|\vec{E}_1'|^2$ and $|\vec{E}_0'|^2$ were given by (6).

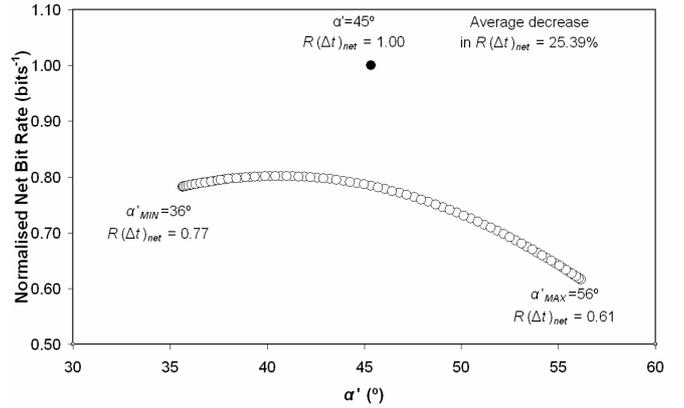

Fig. 11. The effect of PDL on the net bit rate ($R(\Delta t)_{net}$) for a fixed QBER of 3.59%. The solid circle represents $R(\Delta t)_{net}$ for a system with a PDL of 0 dB at a relative angler $\alpha$ of 45˚. The lower curve shows the effect of a PDL of 2.23 dB on $R(\Delta t)_{net}$ assuming an initial relative angle $\alpha$ of 45°.

The average decrease of $R(\Delta t)_{net}$ due to a PDL of 2.23 dB was calculated to be 25.39%. However, as mentioned previously, the $R(\Delta t)_{net}$ represented in Fig. 11 was calculated for a fixed QBER of 3.59%. The average decrease on $R(\Delta t)_{net}$ was also calculated for different values of QBER (%) and is shown by the upper solid line in Fig. 12. The maximum QBER was 9.3%, as for a QBER in excess of this, $R(\Delta t)_{net}$ for a system with a PDL of 2.23 dB falls to zero for certain values of the angles $a$ and $b$.

### F. PDL compensation

Alice and each Bob could use a PDL compensator scheme with a power monitor at each output port of the splitter to reduce the effects of the high PDL on the transmitted states in the point-to-multipoint PON. The power monitor is connected to a feedback compensator which controls an automated polarization controller (APC) in Alice. The power monitor is used to measure the loss incurred by the polarization states after they are transmitted through the element with PDL - the telecomms splitter in this case. If this loss is above a specified threshold, the APC, controlled by the feedback compensator, chooses another two non-orthogonal states at different angles $a$ and $b$. If the loss of these new polarization states is measured to be below a predetermined maximum threshold loss, the signal is transmitted to one user of the network who we shall designate $Bob_i$. Hence Alice sends data using only the polarization states which lead to the minimum PDL for $Bob_i$.

Fig. 12 shows the decrease in $R(\Delta t)_{net}$ for $Bob_i$ when using PDL compensation compared to the case where no PDL compensation is used.

The PDL compensation scheme reduces the decrease in $R(\Delta t)_{net}$ caused by the PDL, especially at high values of the QBER, even to the point where a gain is observed for QBER > 8.5%. For high values of the QBER $R(\Delta t)_{net}$ decreases rapidly with $\alpha'$ due to the term proportional to $I_{AE}$ in (1). Hence $R(\Delta t)_{net}$ for a system with no PDL and $\alpha = \alpha' = 45°$ is lower than $R(\Delta t)_{net}$ for a system with a PDL of 2.23 dB and $\alpha' = 36°$ despite the additional losses on the non-.

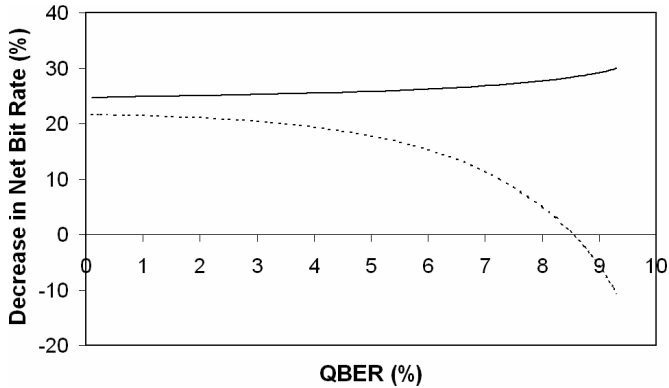

Fig. 12. Decrease in net bit rate for a system with a PDL of 2.23 dB using PDL compensation (lower, dotted line) and with no PDL compensation (upper, solid line).

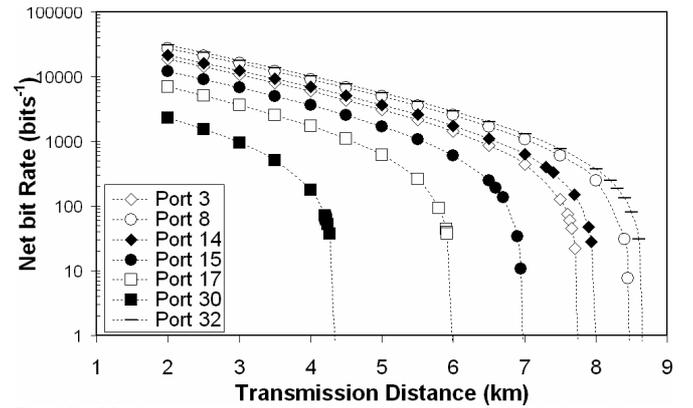

Fig. 13. Maximum transmission distances for seven Bobs in the point-to-multipoint PON.

orthogonal states caused by the PDL.

It is important to note, however, that PDL compensation cannot be performed for all the users of the network simultaneously, since the values of $a$ and $b$ that offer some Bobs the optimal value of $R(\Delta t)_{net}$ are not necessarily favorable angles for others. Those users not using PDL compensation would be subject to a higher average loss in their $R(\Delta t)_{net}$, that depends on the QBER, as shown by the solid line represented in Fig. 12.

### G. Maximum Transmission Distance for Each User of the Point-to-Multipoint PON

It is obvious that the users of the point-to-multipoint PON connected to ports that exhibit higher insertion losses will be more limited in the maximum length of their quantum channel. Utilizing one Bob (at a transmission distance of 2 km in Fig. 8) used for the QBER and $R(\Delta t)_{net}$ measurements as a reference, the QBER and $R(\Delta t)_{net}$ were estimated for the remaining receivers, connected to other ports, using [11]:

$$QBER = \kappa + \frac{D}{2B}, \qquad (8)$$

where $\kappa$ takes into account the error introduced due to the imperfection of all the optical components in the system, $B$ is the bit-rate measured at Bob and $D$ is the dark noise at Bob's detector. Fig. 13 shows the estimated maximum transmission distance at 1 GHz clock frequency for some ports of the telecomms wavelength splitter, including the ports with the highest and lowest insertion loss, ports 30 and 32, respectively. The difference in the maximum transmission distances for these two ports corresponded to 4.3 km. The same technique was applied to the $\lambda \sim 850$ nm splitter and the maximum difference for the transmission distance between ports corresponded to only 0.5 km.

### H. Fraction of Shared Bits between Users

An ideal QKD system would use true single-photon sources that only produce one photon for each driving pulse. However, due to the relative infancy of such devices, most current experimental systems use multi-photon laser pulses that are attenuated to a mean photon number per pulse of less than 1 – so-called *weak coherent pulses* (WCPs).

The photon number distribution of a source emitting WCPs follows a Poissonian distribution so that there is always a probably of multi-photon pulses (i.e. pulses containing 2 or more photons) occurring. Clearly, these multi-photon pulses could split between different Bobs at a passive optical splitter, leading to two, or more, Bobs sharing a fraction of the bits in the final key.

A Hanbury-Brown Twiss experiment [24] was used to measure the fraction of photons shared between two users of the two fiber-based access network architectures studied. The mean photon number at *the input of the splitter* was increased over the range from 0.1 to 3 photons per pulse and the experimental multi-photon rate ($M_e(\mu)$) was recorded using time correlated single-photon counting electronics. Several pairs of ports for each splitter were used for comparison purposes.

The theoretical multi-photon rate, $M_t(\mu)$, was approximated for both splitters using [21]:

$$M_t(\mu) \approx \sum_{i=2}^{N} f \left\{ \frac{1}{4} \left[ 1 - \exp\left( -\frac{1}{N} \beta \eta \mu \right) \right] \right\}^i, \qquad (9)$$

where $f$ is the repetition frequency of the photon source, $N$ is the number of output ports of the splitter, $\beta$ the transmission fraction after loss, $\eta$ the detection efficiency of the detector and $\mu$ the mean photon number per pulse. The factor ¼ is due to the fact that in the B92 protocol, only ¼ of the photons are detected unambiguously.

Figs. 14 and 15 show $M_t(\mu)$ and $M_e(\mu)$ versus the mean photon number per pulse for two output ports of the 1×32 telecomms wavelength splitter used in the point-to-multipoint PON and for two output ports of the 1×8 $\lambda \sim 850$ nm splitter used in the PON P2P. The error bars for $M_t(\mu)$ were calculated assuming the highest insertion loss for each port, which occurs for a maximum value of PDL. The theoretical and experimental values for $M_t(\mu)$ and $M_e(\mu)$ show a high degree of agreement.

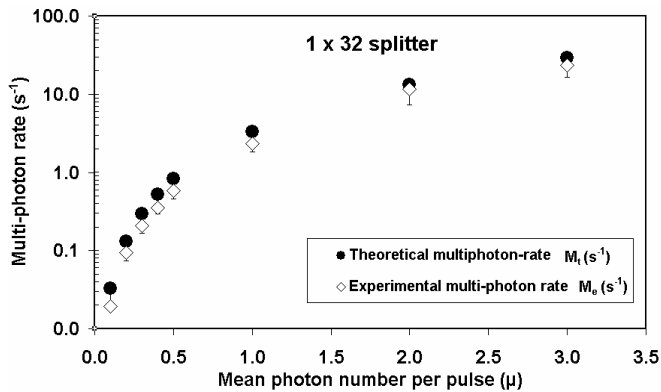

Fig. 14. A graphical plot of the theoretical multi-photon rate for the 1×32 splitter used in the point-to-multipoint PON and the measured experimental values, versus the mean photon number.

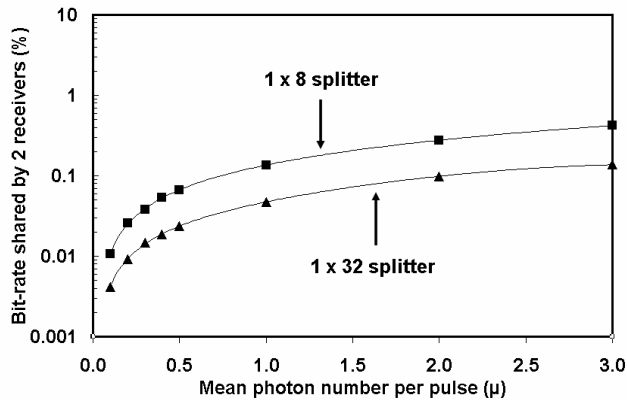

Fig. 16. The shared bit rate for two Bobs for various mean input photon numbers.

Fig. 16 shows the percentage of bits shared (on average) by two users connected to two ports of the 1×32 or the 1×8 optical splitters from the detection rate at any single receiver. For $\mu \sim 0.1$ photons per pulse at the input of the splitter the shared bit rate between two receivers was measured to be an average of 0.042% for the 1×32 telecomms wavelength splitter configuration and 0.011% for the 1×8 $\lambda \sim$ 850 nm splitter. For the latter configuration $\mu$ was also set to $\sim$ 0.8 at the input port, which gave a percentage shared bit rate of 0.11%. The percentage of the bit rate shared between one receiver ($Bob_i$) and the rest of receivers on the network can easily be calculated by multiplying the shared bit rate between two receivers by the number of receivers of the network minus one (i.e. N-1).

$Bob_i$ can reduce this fraction of bit-rate shared with the rest of the receivers of the network to zero by adding an extra step in the standard process of privacy amplification [8]. The first step allows the two legitimate parties to reduce the valid information held by an eavesdropper to zero worth by algorithmic manipulation of their distilled key information. The second step applies the same process to the knowledge shared between different Bobs on the network, allowing them to reduce the information shared between them to zero.

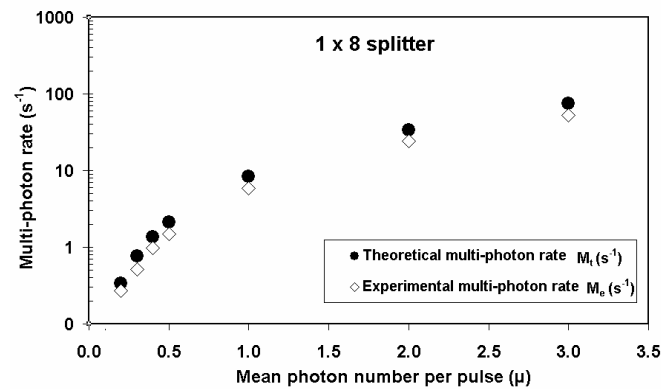

Fig. 15. A graphical plot of the theoretical multi-photon rate for the 1×8 splitter used in the PON P2P and the measured experimental values, versus the mean photon number.

Naturally, with true single-photon sources there is no possibility for the individual data bits to be split at the passive optical splitter, removing this need for an additional step in the privacy amplification.

## IV. CONCLUSION

Two possible multi-user access networks based on those currently being implemented to increase the bandwidth available to customers have been designed and tested for QKD. The first uses a passive optical splitter within the central node, enabling a PON approach to multiple point-to-point links, which has an associated reduction in cost and complexity compared to using a dedicated sender for every receiver. The second multi-user access network is a point-to-multipoint PON which uses a standard telecomms splitter in the transmission channel to randomly route the photons from the sender into the receivers. Excellent performance is achieved for both multi-user QKD architectures, with QBER values < 10% and net bit rates up to 147 kilobits$^{-1}$, depending on configuration and distance.

The first multi-user QKD architecture offers the potential of higher bit-rates as setting the mean photon number per pulse after the splitter compensates for the excess internal losses of the splitter.

In the second multi-user QKD architecture the optical splitter, which is single mode at $\lambda \sim$ 1550 nm, suffers from an increased PDL when used at $\lambda \sim$ 850 nm, which can reduce the net bit-rate. However, the use of PDL compensators between each user and Alice could significantly reduce this decrease, especially for high values of the QBER.

Although the use of weak-coherent pulses in the experiments detailed means that a fraction of the bits transmitted by Alice are shared by two or more Bobs the standard technique of privacy amplification can reduce this shared information to zero value.

ACKNOWLEDGMENT

The authors would like to acknowledge the work of Professor S. D. Cova and Dr I. Rech at the Politecnico di Milano who provided the silicon single-photon detector used in this paper.

**Veronica Fernandez** (M'04) received a BSc. with honors degree (5 years) in physics with electronics from the University of Seville, Spain, in 2002 and the PhD degree in physics at Heriot-Watt University, Edinburgh, UK in 2006.

Dr Fernandez has coauthored several journal articles and conference publication on quantum key distribution. She is a member of the Institute of Physics (UK) and a member of the Institute of Electrical Engineering (UK). In November 2003 she was awarded an IEE Vodafone Scholarship for postgraduate study in telecommunications.

**Robert J. Collins** (M'05) received an MPhys. with honors degree (4 years) in physics from the Heriot-Watt University, Edinburgh, UK, in 2003. In 2004 he started a PhD in physics at Heriot-Watt University.

His areas of research include quantum key distribution, single-photon sources, random number generation, equipment interfacing, data analysis, and software development.

Mr. Collins is a member of the Institute of Physics (UK).

**Karen J. Gordon** received a BSc. degree with honors (4 years) in applied physics and computing from Napier University, Edinburgh, UK, in 1998, the MSc. Degree in optoelectronic and laser devices from St. Andrews University, St. Andrews, UK, in 2000 and the PhD degree in physics from Heriot-Watt University, Edinburgh, UK in 2004.

In 2003 she undertook a postdoctoral position with Heriot-Watt University researching quantum key distribution and related technologies and has coauthored several journal articles, conference publications and patent applications in the area of quantum key distribution.

Dr Gordon is a member of the Institute of Physics (UK) and a member of the Optical Society of America.

**Paul D. Townsend** received the BSc degree in physics from the University of East Anglia, Norwich, UK in 1983 and the PhD degree in physics from the University of Cambridge, Cambridge, UK, in 1987.

From 1987 to 1990 he held a joint position with St. John's College and with Bellcore, Red Bank, NJ. In 1990, he joined BT Laboratories, Ipswich, UK, where he worked on various aspects of quantum optics and optical communications including quantum cryptography. In 2000, he joined the Corning Research Centre, Ipswich, where he was Project Manager for access and metro networks applications research. Since 2003, he has been with the Department of Physics, University College, Cork, Ireland, where he leads the access networks and quantum communications research activities in the Photonics Systems Group.

Prof. Townsend has coauthored more than 100 journal articles, conference papers, and patent applications in the areas of polymer physics, quantum and nonlinear optics and optical communications. He is a Fellow of the Institute of Physics (UK) and an Honorary Professor in the School of Engineering and Physical Sciences at Heriot-Watt University, Edinburgh, UK.

**Gerald S. Buller** (M'06) received the BSc degree (with honors) in natural philosophy from the University of Glasgow, Glasgow, UK, in 1986 and the PhD degree in physics from Heriot-Watt University, Edinburgh, UK, in 1989.

He was made a professor of physics at Heriot-Watt University in October 2006. Since 1990, he has led the Photon-Counting Group in researching time-resolved photoluminescence, time-of-flight ranging and quantum key distribution and has coauthored more than 200 journal articles, conference papers, and patent applications in the areas of photon-counting, semiconductor optoelectronics, optical coatings, and optical interconnects.

In 2002 Prof. Buller founded Helia Photonics Ltd. of Livingston, Scotland. He is a Fellow of the Institute of Physics (UK) and a member of the Optical Society of America.